# Exploring the phases of 3D artificial spin ice: From Coulomb phase to magnetic monopole crystal

M. Saccone[1*], A. Van den Berg[2*], E. Harding[2], S. Singh[2], S. R. Giblin[2], F. Flicker[2] and S. Ladak[2]

1. Center for Nonlinear Studies and Theoretical Division, Los Alamos National Laboratory, Los Alamos, NM, USA.

2. School of Physics and Astronomy, Cardiff University, The Parade, Cardiff, United Kingdom CF24 3AA

* Authors contributed equally.

**Artificial spin-ices consist of lithographic arrays of single-domain magnetic nanowires organised into frustrated lattices. These geometries are usually two-dimensional, allowing a direct exploration of physics associated with frustration, topology and emergence. Recently, three-dimensional geometries have been realised, in which transport of emergent monopoles can be directly visualised upon the surface. Here we carry out an exploration of the three-dimensional artificial spin-ice phase diagram, whereby dipoles are placed within a diamond-bond lattice geometry. We find a rich phase diagram, consisting of a double-charged monopole crystal, a single-charged monopole crystal and conventional spin-ice with pinch points associated with a Coulomb phase. In our experimental demagnetised systems, broken symmetry forces formation of ferromagnetic stripes upon the surface, a configuration that forbids the formation of the lower energy double-charged monopole crystal. Instead, we observe crystallites of single magnetic charge, superimposed upon an ice background. The crystallites are found to form due to the intricate distribution of magnetic charge around a three-dimensional nanostructured vertex, which locally favours monopole formation. Our work suggests that engineered surface energetics can be used to tune the ground state of experimental three-dimensional ASI systems.**

The many-body interaction of dipoles is crucial to understanding a diverse range of phenomena across physics, with its long-range anisotropic nature yielding a wealth of fascinating phenomena. For example, dipolar interactions can yield novel vortex stripes in an ultracold quantum gas[1], a low temperature residual entropy in frustrated condensed matter systems[2] and Rosensweig instabilities in ferrofluids[3], yielding self-organised surface structures. The pioneering work of Luttinger and Tisza[4] provided a foundation for understanding dipolar ordering in simple lattice geometries, but this was only extended recently to arbitrary geometries[5]. To date, the experimental placement of dipoles into complex 3D arrangements has been lacking, with scientists mainly relying upon arrangements provided by condensed matter systems. One such model system, known as spin-ice[6], has been studied intensively and has allowed systematic study of frustration and associated emergence[7]. These systems consist of rare earth moments on corner sharing tetrahedra. The Hamiltonian consists of dipolar and exchange terms and the ~10 Bohr magneton moment means that dipolar interactions are important in determining the nature of the ground state. Since all pairwise interactions within a single tetrahedra cannot be simultaneously satisfied, the system is geometrically frustrated. This yields a local ordering principle known as the ice-rule, in which two spins point into the centre of a tetrahedron and two spins point out, yielding a macroscopically degenerate ground state and a residual entropy measured at low temperature. Interestingly, Monte-Carlo (MC) simulations which encompass sufficient dynamics via a loop algorithm find an ordered phase in spin-ice at very low temperatures, which consists of stripes of anti-parallel spins[8] but so far this has not been measured experimentally.

A new framework to understand the physics of spin-ice was later proposed which treated each spin as a dimer consisting of equal and opposite charges[9,10]. Within this framework local excitations above the ice-manifold are magnetic monopoles in the vector fields *M* and *H*, since once the chemical potential has been surpassed, they interact via a magnetic equivalent of Coulomb's law. Subsequent studies provided experimental evidence of magnetic charge transport in bulk spin-ice materials[11,12]. The ground state of spin-ice and the associated dynamic route can then be considered within the framework of magnetic charge, where the ratio of the chemical potential to the magnetic Coulomb energy of a nucleated pair is an important quantity[13]. When this effective chemical potential approaches a value of half the Madelung constant ($M/2 = 0.819$ for a diamond lattice), a magnetic charge crystal is expected, whereby charges of alternating polarity are tiled throughout the structure[13]. For the canonical spin-ice materials, the effective chemical potential is 1.42, suggesting the monopoles are free to propagate through the system, yielding a disordered spin-ice phase. To observe charge-ordered states in bulk solid-state systems, one needs to find systems with specific material properties. One example, the spin-ice candidate $Nd_2Zr_2O_7$ has recently shown charge crystal behaviour, combined with disordered spin background, a signature of magnetic fragmentation whereby the local magnetic moment splits into divergence-full and divergence-free parts[14]. A tuneable, engineered system has the capability to explore this phase space systematically.

Artificial spin-ice materials are arrays of lithographically patterned single-domain nanomagnets[15,16]. As such they are a powerful means to explore ordering in dipolar systems by design. Initial studies focussed upon simple square[15] and Kagome arrays[17], which has subsequently been extended to a wide range of 2D geometries providing a means to explore a variety of model spin systems in statistical physics and more exotic phenomenon such as topological frustration in the Shatki lattice[18] and superferromagnetism in pinwheel lattices[19]. To date, most ASI studies have focussed upon 2D systems due to ease of fabrication but interest has spanned into layered systems[20,21] with both theoretical and experimental studies investigating how these can be used to realise a range of ground states including model vertex systems[22] and superlattice structures[23].

The advent of three-dimensional lithography now allows the creation of lattices that directly mimic bulk spin-ice geometries[24-26], but with tunability to control factors such as magnetic moment and lattice spacing. Such 3D artificial spin-ice (3DASI) systems, which have a Hamiltonian governed purely by dipolar energetics have only recently been experimentally realised and through simple linear field driving protocols, magnetic charge has been directly observed across the 3DASI surface[24].

In this article we first use finite temperature MC simulations to carry out a detailed mapping of ordering in idealised 3DASI systems within a diamond-bond lattice geometry. We find a rich phase diagram consisting of a double-charged monopole crystal, single-charged monopole crystal and a spin-ice phase. We move on to measure the demagnetised state in an experimental 3DASI system and find evidence of an out-of-equilibrium state, whereby crystallites of magnetic charge are superimposed upon an ice background.

**Simulating the phase diagram of an idealised 3D artificial spin-ice**

**Figure 1a** and **1b** shows a schematic of the simulated unit-cell geometry. Compass needle dipoles are placed upon a diamond-bond lattice, which has a lateral extent of 15 x 15 unit cells and a thickness of a single unit cell. To aid in discussion, we define a series of sub-lattices which are labelled L1 – L4. The upper surface terminates in coordination-two vertices (L1), below which two layers of coordination-four vertices are found (L2, L3). Finally, the lower lattice surface again terminates in coordination-two vertices (L4). This geometry matches our experimental 3DASI system. The compass needle model (see Methods), is equivalent to treating magnetic dipoles as two magnetic monopoles with a small variable separation. We use a metropolis algorithm to determine the ground state of the system as a function of the dipole length (b), with a fixed lattice spacing (a=1), over a range of temperatures. **Figure 1c** and **1d** show an overview of the phase diagram as a function of *b* over a range of temperatures, whilst **Figs 1e** and **1f** show the specific heat $C_v$ and corresponding entropy per site s for four values of b. To facilitate interpretation, we define an order parameter ($M_c$, see Methods) which quantifies the extent to which a magnetic charge crystal has formed. For lower temperatures, a high *b* lattice yields strong local Coulomb interactions upon vertices, forcing charge neutrality and a spin ice ground state as can be seen in **Fig 1c** and **1d**. A representative arrow map of the spin-ice state is shown in **Fig 1g**. Ice vertices dominate the microstate occurring at frequencies reflecting underlying vertex probabilities (ergodic balance). The surface L1 layer forms short ferromagnetic strings as seen in previous theoretical studies[27]. The magnetic structure factor (**Fig 1h**) shows pinch points associated with a Coulomb phase and signatures of short-range magnetic strings with diagonal lines seen along **q**=[1,1] and **q**=[-1,-1]. At b=1, low temperature, the ground state entropy $s_0$ of spin ice is evident (**Fig 1f**). In the Methods we calculate $s_0$ analytically using two models: first, using Pauling's method of independent tetrahedra which is well tested in bulk spin ice. Second, by assuming that the surfaces order first and constrain subsequent layers. **Fig 1f** shows a closer agreement with the latter model, a fundamental difference between the bulk and slab geometries. As b decreases, the frustration and ground state entropy disappear.

Reducing b lowers the chemical potential and in the low temperature regime this yields a phase transition to a double charge crystal (CII). Of particular interest is how such a crystal forms whilst constrained to an odd number of charge layers. The state is characterised by ±2q charges upon surface coordination-two vertices (L1 and L4) and ±4q charges upon L1/L2 and L3/L4 coordination-four vertices, as portrayed in **Fig 1i** and **Fig 1d**. The order parameter ($M_c$) of this CII state is found to be greater than 0.8, as seen by the yellow region in **Fig 1c**. A neutral layer is found in the centre, consisting of type I vertices. Notably, the sheet geometry produces a coarse-grained field that is approximately constant with respect to distance. This makes the inclusion of a neutral spacing layer more negligible. The magnetic structure factor (**Fig 1j**) shows clear Bragg peaks due to antiferromagnetic order and associated charge ordering. With intermediate values of b, and at higher temperature, one of the coordination-four double-charged sheets "spreads" into the neutral layer, creating two consecutive single charge sheets, as depicted in **Fig 1k** and a cross-sectional view shown in **Fig 1d**, right-panel. This state is named CI. This increases the entropy of the system while maintaining a relatively favourable environment for charges. As temperature is further increased, a peak in specific heat (**Fig 1e**), corresponding increase in entropy (**Fig 1f**) and decrease in $M_c$ (**Fig 1c**) indicates a phase transition to a paramagnetic state. Overall, the phase diagram described by MC simulations is also captured analytically with a simple mean field analysis (See Methods).

**Exploring the ordering in experimental 3D artificial spin-ice systems**

A 3DASI system was fabricated to explore the extent to which the idealised theoretical phase diagram can be captured experimentally. The system was fabricated using a combination of two-photon lithography and evaporation (See Methods) [24,28]. **Figure 2a** shows a scanning electron microscopy (SEM) image of the array which takes a diamond bond lattice geometry and has a lateral extent of 50 μm x 50 μm. **Figure 2b** shows a zoomed top-view, false-colour SEM image with the upper four sub-lattices labelled (L1 – L4). As in the simulated systems, the lattice terminates in coordination-two vertices upon the surface, with typical coordination four vertices found below at the L1/L2 and L2/L3 junctions. The lower L4 sub-lattice, again terminates in coordination-two vertices.

Our previous work has shown that individual nanowires are single domain and magnetic force microscopy (MFM) can be used to determine the contrast for different vertex types[24]. We now exploit this to determine the demagnetised configuration obtained in 3DASI systems. Note, due to the limited resolution of MFM with lift height, we are only able to measure contrast upon the upper three layers, L1-L3. MFM was performed over large portions of the lattice after planar demagnetisation protocols (See Methods). All vertex types observed in previous experiments[24], including ice-rule vertices with zero magnetic charge and monopole states with magnetic charge $Q=\pm 2q$ are again observed (**Fig 2d**). The demagnetised array also contains previously unseen monopole states of charge $Q=\pm 4q$, as can be seen in **Fig 2e**.

**Figure 3a** shows an experimental magnetic charge map of a 30μm x 30μm region of the lattice, determined by MFM. Three distinct phases are measured and can be readily identified in the charge map with detailed configuration shown in **Figs 3b-d**. Magnetic charge crystallites can be seen with ±2q tiling, as highlighted by the green box in **Fig 3a**. An arrow map of a typical charge crystallite region is shown in **Fig 3b**, which shows that it arises due to two types of distinct ordering. The L1 sub-lattice that consists of alternating coordination two and coordination four vertices is found to order into ferromagnetic stripes. Analysis of the L1 sub-lattice **(Extended Data Fig 1)**, shows that this is the case for the entire measured area, with coordination-two monopoles being very rare and only observed upon <1% of vertices consistent with previous work[24]. Over large regions of the measured area (~20 %), including in the charge crystallite regions, the L2 sub-lattice is found to host anti-ferromagnetic ordering (**Extended Data Fig 2a,b**). Breaks in the antiferromagnetic ordering upon L2, via short ferromagnetic strings occurs frequently, with frequency decaying with string length (**Extended Data Fig 2c**). Interestingly, we find that breaks in the antiferromagnetic ordering often occurs to mitigate the formation of +/- 4q charges. We note that since the configuration of the charge crystallites observed experimentally (CI$_E$) has ferromagnetic stripes on L1, it is distinct to the CI charge crystal seen in simulations.

Between areas of magnetic charge crystallite, large patches of the ice phase are observed, as shown by the orange region in **Fig 3a**, with full representative arrow map shown in **Fig 3c**. These ice regions are largely composed of type II vertices, which due to a subtle broken symmetry in 3D geometry, are the lowest energy vertex type according to micro-magnetic simulations[24]. Finally, only very small regions of the double-charge (CII) crystallite are observed as shown by purple region in **Fig 3a** and associated arrow map in **Fig 3d**. The full measured region is shown in **Extended Data Fig 3a**, with associated vertex types shown in **Extended Data Fig 3b** and vertex charge shown in **Extended Data Fig 3c**. The vertex statistics show a strong preference for type III vertices (61.2%), followed by type II vertices (29.8%). Both low energy type I vertices and high energy type IV vertices are only observed occasionally at 5.3% and 3.6% respectively. As would be expected, our measurements indicate charge neutrality, within error as shown in **Extended Data Fig 3c**. Overall, the charge order parameter as calculated for simulations takes a value of 0.31, for this experimental system (See Methods).

The magnetic structure factor of the entire measured data is shown in **Fig 4a**, with sub-sets corresponding to individual sub-lattices shown in **Figs 4(b-d)**. Focussing upon the data for all layers (**Fig 4a**), the presence of intense Bragg peaks can be seen, superimposed upon weaker diagonal lines along **q**=[1,1]. In order to further interpret this data, we deconvolve the layers. The L1 structure factor (**Fig 4b**) consists of a peak upon **q**=[0,0], indicative of ferromagnetic order on the surface. Weaker split peaks about **q**=[1/2,1/2] come about due to presence of longer period domains upon L1, as demonstrated in **Extended Data Fig 4**. The L2 structure factor (**Fig 4c**) shows peaks due to both type II tiling as well as the magnetic charge crystallite regions as demonstrated in **Extended Data Fig 5**. Finally, the L3 structure factor (**Fig 4d**) shows a diffuse signal, with weak Bragg peaks superimposed. This is consistent with the full arrow map (**Extended Data Fig 3a**), which shows a mixture of charge-ordered and ice states upon L3. Further breakdown of the structure factor via layer and region can be found in **Extended Data Fig 6**.

**Magnetic charge crystallite formation**

We now discuss the observed experimental configuration in terms of the states predicted by MC simulations. For the real experimental systems studied here, the scaled needle length (b) depends upon the vertex type (**Extended Data Fig 7**), due to the presence of domain walls close to the vertex. When considering all ice-rule vertices, an average b of 0.89 is obtained, suggesting a Q=±4q monopole crystal would be expected as the ground state. However, a set of Q=±2q crystallites form, superimposed upon an ice background. A number of factors may account for this discrepancy. Previous work has suggested that in experimental 3DASI systems, magnetic charges upon surface coordination two vertices are very unfavourable with micromagnetic calculations of single vertices suggesting such excitations cost a factor of three larger than coordination four monopoles[24]. The immediate implication of this is that ferromagnetic stripes upon L1 will forbid the formation of a Q=±4q monopole crystal, apart from regions with local disorder. This is reinforced by the deterministic demagnetisation protocol which favours the formation of ferromagnetic stripes upon the surface. Given this constraint, the system can only form a single charge crystal. However, the formation of charge crystallites via a demagnetisation routine remains surprising and has not been seen previously in either pristine, traditional 2DASI or more exotic layered 2.5D systems. In the former case, charge crystals can be formed in modified square ASI by utilising an MFM tip to selectively switch islands[29] but demagnetisation of conventional square systems yields a low magnetisation, disordered ice phase with low frequencies of monopole excitations[15]. For pristine Kagome systems, demagnetisation yields a 2-in/1-out ice-rule throughout the lattice[17] with only thermally annealed systems yielding some degree of charge ordering[30]. Modifications of the Kagome geometry, either by tuning island lengths within a single unit cell[31], or by placing exotic nano-bridges at vertices, can also yield charge ordering[32].

Considering the dynamics of the demagnetising protocol and starting in saturating fields, the system becomes uniformly tiled in type II vertices. Though these are the lowest energy state for single vertices[24], the net magnetisation makes these less favourable globally. The effective chemical potential upon L1[24] as modified by surface energetics ($\mu^*$=1.22) means that deconfined monopoles nucleate and propagate for each 180 degree rotation of the field. At threshold fields, nucleation events upon L1 become less likely, leaving long ferromagnetic strings as observed in the experimental data. The effective chemical potential upon the L2 sub-lattice[24], within a simple dipolar approximation is lower ($\mu^*$=1.03) and favours the local production of correlated charge pairs (type III vertices). It is interesting to note that type III vertices also have a slightly lower b value (0.78) due to the complex 3D distribution of magnetic charge around the vertex. When taking into account the reduced separation between nucleated charges, this reduces the effective chemical potential and yields (See methods) a value of $\mu^*=0.91$, approaching the critical value of $M/2=0.819$. The implication of this is that it is very favourable for a monopole pair to nucleate and remain correlated. Once a single pair is formed, local vertex-vertex Coulomb interactions are minimised by tiling charges of opposing sign. The residual ice-rule regions reflect regions which have not yet equilibrated. It is possible that longer or more complex 3D demagnetisation protocols will promote more efficient exploration of the energy landscape, allowing such ice regions to be further minimised. Altogether, returning artificial spin ice to its three-dimensional origins unlocks previously inaccessible exploration of phase space. Not only is the vertex symmetry restored, but the charge excitations are brought close enough together to encourage never before seen charge crystallite formation from a charge neutral background. We anticipate that fine control of 3D vertex geometry and NiFe thickness will allow suppression of surface energetics and together with an exploration of more complex demagnetisation protocols, or thermal relaxation will allow a realisation of the double charged crystal. It is also expected that more sophisticated synchrotron techniques[35] may also allow imaging of systems greater than one-unit cell in thickness.

**Acknowledgments**
SL gratefully acknowledges funding from the Engineering and Physics Research Council (EP/L006669/1, EP/R009147) and Leverhulme Trust (RPG-2021-139). SRG acknowledges funding from the EPSRC (EP/S016554/1) and Leverhulme Trust (RPG-2021-139). The work of M.S. was carried out under the NNSA of the U.S., DoE at LANL, Contract No. DE-AC52-06NA25396 (LDRD grant - PRD20190195).

**Author Contributions**
SL conceived of the study, supervised experimental work and wrote the first draft of the manuscript. AV carried out sample fabrication, magnetic force microscopy (MFM), micromagnetic simulations and analysed experimental data. EH carried out MFM, analysed the experimental data and analysed micromagnetic simulation data. MS wrote code to carry out the Monte-Carlo simulations, derived the mean field theory and with AV, performed simulations to determine the phase diagram of 3D artificial spin-ice. SS wrote the code to calculate magnetic structure factors and together with AV, SG, FF and SL, interpreted the data. FF carried out entropy calculations of the 3DASI system in paramagnetic and spin-ice phases. All authors contributed to the writing of the final manuscript.

**Data Availability**
Information on the data presented here, including how to access them, can be found in
the Cardiff University data catalogue.

**Methods**

**Entropy Calculations**

In this section we provide the details of our analytical calculation of the ground state entropy of our spin ice model. We employ the following conventions.
- We adopt the graph theory terminology of vertices connected by edges. In our system each edge hosts a single, Ising-like magnetisation.
- Two edges meet at each vertex in layers $L_1$ and $L_5$. Four edges meet at each vertex in layers $L_2$, $L_3$, $L_4$ ; each of these vertices shares two edges with vertices in each of its neighbouring layers.
- There are the same number of vertices, $N$, in each layer.
- Let $S_n$ be the entropy in layer $n$, in units where $K_B=1$, and $s_n=S_n/N$ *is the entropy per vertex in layer $n$*.

- Let *S* be the total entropy of the system, and *s=S/(5N)*.

***Paramagnet***: we use the high-temperature paramagnetic phase to constrain the entropy in our numerical calculations. Each domain's orientation is independent. In layers 1 and 5 there are two edges per vertex, giving $s_{1,5}$ = *log $2^2$* , and in the other layers there are 4 edges per vertex giving $s_{2,3,4}$ = *log $2^4$* . Therefore,

$$s = (16/5)\log(2) \approx 2.22 \tag{1}$$

***Spin ice***: for the ice state without field annealing (the physically relevant case around *b/a ≈ 1*) we have a net charge of zero at every vertex. Each $L_1$ vertex therefore has a net magnetization (1-in 1-out means both domains align). It appears the $L_1$ vertices are completely uncorrelated, even along a single $L_1$ line. Therefore, there are 2 choices per vertex, and $s_1$ = $s_5$ = *(1/5) log(2)* . Each $L_2$ vertex now has two of its domains fixed by $L_1$ vertices, giving no freedom in these two domains. There are two remaining independent choices per vertex, giving $s_{2,4}$ = *(1/5)log(2)* . $L_3$ is then completely constrained by $L_2$ and $L_4$ . Therefore $S_3$ =0 . Overall,

$$s = (4/5)\log(2) \approx 0.55 \tag{2}$$

This agrees with our numerically calculated result to within standard error. The value differs from the Pauling estimate in bulk spin ice; this is because surface energetics dominate in our single-unit-cell slabs.

*Monte Carlo Simulations*

The interaction energy between two artificial nanomagnets may accurately account for their finite size through the compass needle model. That is, the energy, $E_{ij}$, between magnets *i* and *j* is approximated by considering two point charges at the end of each nanomagnet that interact with Coulomb attraction or repulsion:

$$E_{ij} = \alpha_{ij} \frac{\mu_0 m^2}{4\pi L^2} \left[ \frac{1}{|\boldsymbol{r}_{ai}-\boldsymbol{r}_{aj}|} - \frac{1}{|\boldsymbol{r}_{ai}-\boldsymbol{r}_{bj}|} - \frac{1}{|\boldsymbol{r}_{bi}-\boldsymbol{r}_{aj}|} + \frac{1}{|\boldsymbol{r}_{bi}-\boldsymbol{r}_{bj}|} \right] \tag{3}$$

Here $\mu_0$ is the permeability of free space, $m$ is the nanomagnet's magnetic moment, and $L$ is the nanomagnet length. $\boldsymbol{r}_{ai}$ is the position of a magnetic charge, the first index $a$ referring to it being positive and the second index $i$ denoting to which magnet it belongs. $\alpha_{ij}$ is the surface energy factor, which for data presented in this publication was set to 1. Since nanomagnet length wildly influences energy scales, all computational energies were normalized by their strongest interactions, such that $\tilde{E}_{ij} = E_{ij}/E_{max}$ .

From this energy we can see that increasing length of the magnets increases nearest neighbour dominance. It's worth noting that the exact distribution of charge and, therefore, precisely what the "length" of the magnets is, depends largely on the details of the nanomagnet's geometry and domain wall arrangement. In this study, this energy is used in the evaluation of a metropolis method Monte Carlo analysis.

*Effective Chemical Potential Calculations*

The chemical potential of a coordination-four vertex, upon a diamond bond geometry has previously been calculated within a dipolar framework. The energy between any pair of dipoles can be written as:

$$E_{12} = \frac{u}{4} \frac{|\hat{m}_1 \cdot \hat{m}_2 - 3(\hat{m}_1 \cdot \hat{r})(\hat{m}_2 \cdot \hat{r})|}{\left|\frac{r}{a}\right|^3} \quad (4)$$

Where m represents the magnetic moment unit vector, r is the moment separation, a is the lattice constant and u is the Coulomb energy between charges:

$$u = \frac{\mu_0 Q^2}{4\pi a} \quad (5)$$

with $Q = 2m/a$. One can then simply write the chemical potential as the energy difference between a monopole and an ice-rule state, offset by the magnetic Coulomb interaction, between created charges:

$$\mu = (E_{monopole} - E_{lr} - E_{ice}) \quad (6)$$

with

$$E_{lr} = \frac{u}{\frac{r_{charge}}{a}} \quad (7)$$

Assuming a perfect dipolar model whereby the charges are separated by a single lattice spacing yields a chemical potential $\mu$ of 1.03u. The effective chemical potential is therefore $\mu^* = \frac{\mu}{u} = 1.03$

However, in our real experimental system the charge separation in the monopole state is reduced, with $r_{charge} \approx 0.8a$, yielding a reduced $\mu^* = 0.91$. This locally promotes the formation of charge crystallites.

*Magnetic Charge Crystal Order Parameter*
In charge ordered systems, twofold degenerate patterns emerge as the ground states. To measure similarity to these states, we can calculate a charge crystal order parameter defined as:

$$M_c = |\sum_i \Delta_i Q_i| \quad (8)$$

$\Delta_i$ is a template of +1's, -1's, and 0's representing a ground state. This was used to calculate the order parameter for both Monte Carlo simulations, as shown in Fig 1c and for experiments.

*Magnetic Structure Factor*
In the canonical spin ice materials, spin-flip neutron scattering[12] provides what is probably the clearest evidence of spin ice behaviour. Neutron scattering probes the magnetic structure factor projected along the direction of neutron propagation. In artificial spin ice there is a similar tradition of calculating the structure factors, although neutron scattering is not used as a probe. Instead, the structure factor can be inferred directly by Fourier transforming the MFM image[16,33]. In this work we calculated the magnetic structure factor for spin configurations modelling those in our real lattices, as well as those generated in our Monte Carlo simulations. We calculated the full 3D structure factors before taking the $q_z$ = 0 slice, suitable for modelling what would be seen when Fourier transforming a surface MFM arrow map.

*Mean Field Analysis*

Considering the system in the dumbbell model approximation,

$$E = \frac{1}{2} \sum_{i \neq j} K_{ij} Q_i Q_j + \mu \sum_i Q_i^2 \tag{9}$$

where $Q_i = \pm 2, \pm 1, 0$ is the value of the charge on the $i$th vertex, $K_{ij}$ is the interaction strength between charges, and $\mu$ is the chemical potential of a charge. We can calculate the Maxwell-Boltzmann distribution in the mean field approximation and motivate how charge ordering differs from spin ice ground states. Taking the change of variables $Q_i = \Delta_i X_i$, where $\Delta_i$ is a general charge ordered ground state, and introducing a perturbative "field" to this variable $h$ which will later be set to zero,

$$E = \frac{1}{2} \sum_{ij} K_{ij} \Delta_i \Delta_j X_i X_j + \mu \sum_i X_i^2 - h \sum_i X_i. \tag{10}$$

The variable is approximated by deviations from its mean value, $X_i = \langle X \rangle + \delta X_i$. The energy gained by a charge ordered state is called the Madelung constant, which can be written as $\alpha = -\frac{1}{N} \sum_{ij} K_{ij} \Delta_i \Delta_j$ and $\alpha = -\sum_j K_{ij} \Delta_i \Delta_j$. Substituting then yields

$$E = N \langle X \rangle^2 \left(\frac{\alpha}{2} - \mu\right) + [(-\alpha + 2\mu)\langle X \rangle - h] \sum_i X_i \tag{11}$$

from this we can calculate the partition function of a single variable and, because they are independent, $Z = (Z_1)^N$. For a pyrochlore lattice, $Q_i = \pm 2$ ($\Omega = 1$), $\pm 1$ ($\Omega = 4$), and $0$ ($\Omega = 6$) where $\Omega$ is the degeneracy. Substituting $k = \beta(-\alpha + 2\mu)$,

$$Z = \exp\left[-N\langle X \rangle^2 \frac{k}{2}\right] \{ 2\cosh[2k\langle X \rangle - 2h\beta] + 8\cosh[k\langle X \rangle - h\beta] + 6 \} \tag{12}$$

the expectation value of the charge ordering variable is then obtained self consistently from the partition function:

$$\langle X \rangle = -2\tanh\left(k \frac{\langle X \rangle}{2}\right) \tag{13}$$

This self-consistency equation is relatively standard for mean field theories. At high values of $-k$, effectively equivalent to low temperatures, $\langle X \rangle = \pm 2$. As $-k$ becomes closer to zero, these values gradually drop until the system is no longer ordered. This ordering transition may be characterized for small $k\langle X \rangle$ through the first term of the Taylor series:

$$\langle X \rangle = -k \langle X \rangle \tag{14}$$

This is true when $\langle X \rangle = 0$ and $k \leq -1$, meaning below a critical temperature, the system will transition to a nonzero order parameter, corresponding to a charge crystal. That critical temperature is

$$T_c = \frac{\alpha - 2\mu}{k_B} \tag{15}$$

This agrees with the previous experimental results that found a spin ice ground state in systems with a reduced chemical potential greater than $\frac{\alpha}{2}$ and a lack of discrete transition in this regime. The critical temperature also decreases with chemical potential as previously observed. Also, since as temperature approaches zero, the order parameter approaches 2, a double charged crystal is the anticipated ground state. One can justify this by considering the lower entropy of the doubly charged state. Since the experimental system is limited to single charges on the surface, the maximum order parameter we predict for the charge crystal ground state of the pyrochlore thin film with 5 charge sites is $M_c^s = 1.5$.

*Fabrication of 3DASI lattices*
Three-dimensional artificial spin ice lattices were fabricated using two-photon lithography followed by thermal evaporation of $Ni_{81}Fe_{19}$. The coverslips were cleaned in acetone in an ultrasonic cleaner and then washed by isopropyl alcohol (IPA), after which samples were gently dried using compressed air. The coverslip was prepared for TPL with a droplet of immersion oil on one side and Nanoscribe negative-tone photoresist (IPL-780) on the reverse side. The coverslip was then loaded into a Nanoscribe TPL apparatus, and a fabrication script created a number of diamond-bond lattice geometries, each with varying power and scan speed settings. The dimensions of each created lattice are 50μm x 50μm x 10 μm. The completed sample was developed in propyl glycol monomethyl ether acetate (PGMEA) and then rinsed in IPA. An air gun was then again used to remove excess IPA. The sample was then subject to a 50 nm $Ni_{81}Fe_{19}$ evaporation, at a base pressure of 1 x $10^{-6}$ mBar. Approximately 0.06g of evaporated permalloy was used to achieve this thickness based on previous depositions. A crystal quartz monitor (QCM) present during evaporation measured the deposited thickness; this was later confirmed with atomic force microscopy measurements. The resultant structure has a diamond bond geometry polymer scaffold with magnetic material upon the upper surface of nanowires forming a crescent shaped cross-sectional geometry. Due to line-of-sight deposition, the magnetic coating creates a 3DASI lattice which is one unit cell in thickness, as described previously[28]. Individual nanowires are single domain and have a crescent-shaped cross-section with effective width of 200nm and length of 866nm.

*Experimental demagnetisation of lattices*
We used a demagnetising protocol akin to method 1 in a previous publication[34] with a sample rotating at ~1000 revolutions per minute, with axis perpendicular to substrate plane. This effectively yields a rotating magnetic field in the substrate plane. The magnetic field starts at 0 mT and ramps up to 75 mT at 2.5 Ts$^{-1}$ where it is held for 1 second. After this, the field ramps down at 2.5 T s$^{-1}$ to $-75$ mT and is held for a second. The field then oscillates, whilst the magnitude decreases stepwise to zero over a period of five days.

*Magnetic force microscopy (MFM)*

MFM data was captured using a Bruker (Dimension Icon) scanning probe microscope in tapping mode. Ultra-low moment probes were magnetised along the tip axis using a 0.5T permanent magnet. The samples were placed with the L1 sublattice parallel to the probe cantilever with a 45-degree scan angle to the L1 sublattice. MFM data were captured using a 65nm lift-height. Separate scans with reversed tip magnetisation were performed to verify consistency of the contrast, and separate scans with the sample rotated 180 degrees were performed to control for artefacts in the scans.

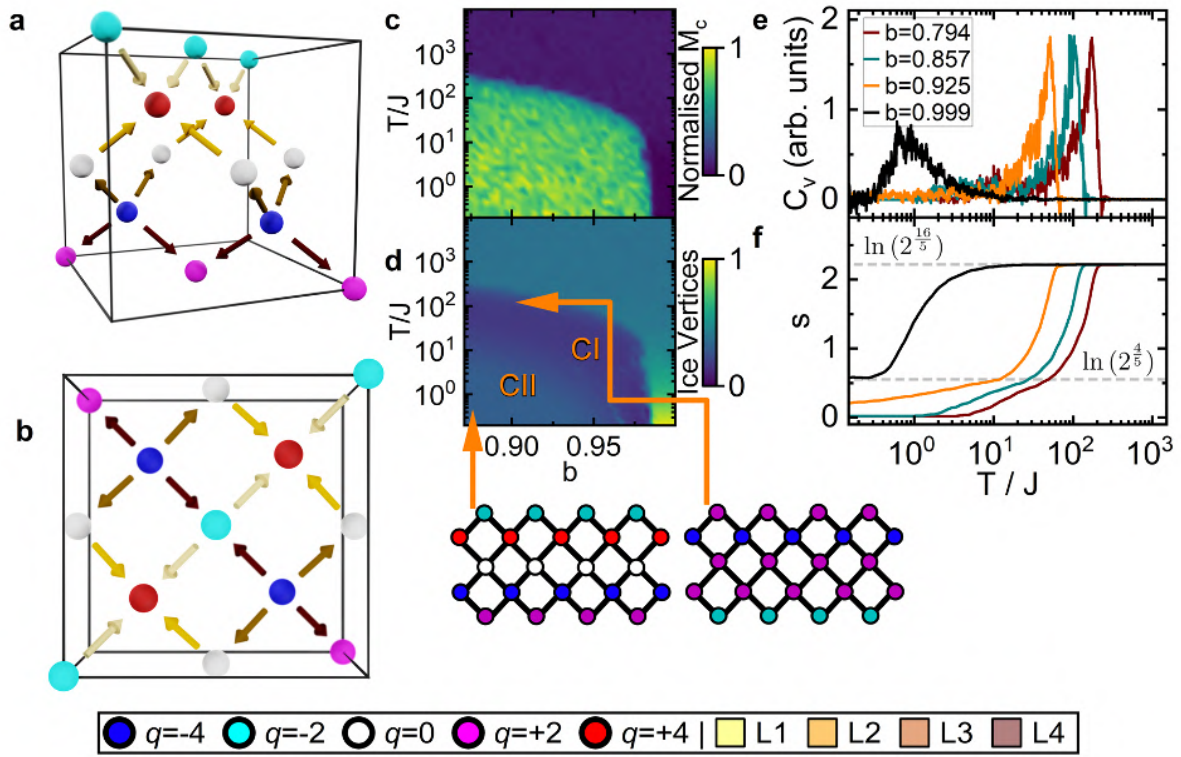

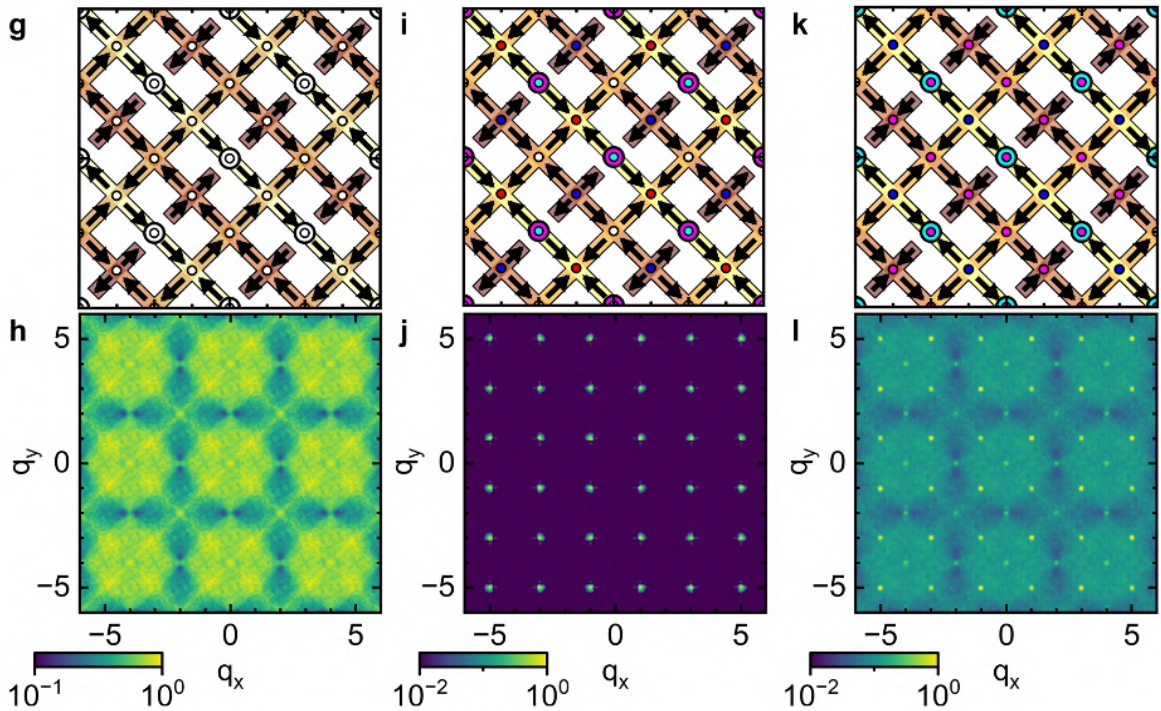

**Figure 1: Simulating the phase diagram in a 3D artificial spin-ice. (a)** A unit cell of the simulated geometry. Spins are placed onto the bonds of a diamond-bond lattice. Magnetic charges are represented as spheres of different colour according to legend. **(b)** View of unit-cell along [001] direction. Arrows are coloured according to layer with cream denoting the surface termination (L1) with coordination-two vertices, yellow (L2), brown (L3) denoting coordination-four vertices and dark red denoting lower surface termination (L4) with coordination-two vertices. **(c)** Phase diagram of 3D artificial spin-ice showing charge crystal order parameter ($M_c$) as a function of
11

reduced dipole length, b and temperature, T. **(d)** The phase diagram now showing variation of ice-rule vertices with reduced dipole length, b and temperature, T. Overall, three main phases are observed. **(e)** The specific heat $C_v$ and **(f)** the entropy per site for four b values with a=1. Entropies were normalised to the high-temperature paramagnetic value. At low temperature for b=1 the residual entropy matches the analytical prediction (see Methods). **(g)** The ice phase, observed for high b and low to intermediate temperatures. **(h)** The magnetic structure factor of the ice phase, showing the typical pinch points associated with a Coulomb phase. **(i)** The double-charge monopole crystal, consisting of ±2q upon the surface terminations, ±4q at L1/L2 and L3/L4 junctions with a neutral layer at L2/L3 vertices. **(j)** The magnetic structure factor associated with the double-charge monopole crystal. **(k)** Phase observed for lower b and intermediate temperatures, showing a single-charged monopole crystal and **(l)** its associated structure factor.

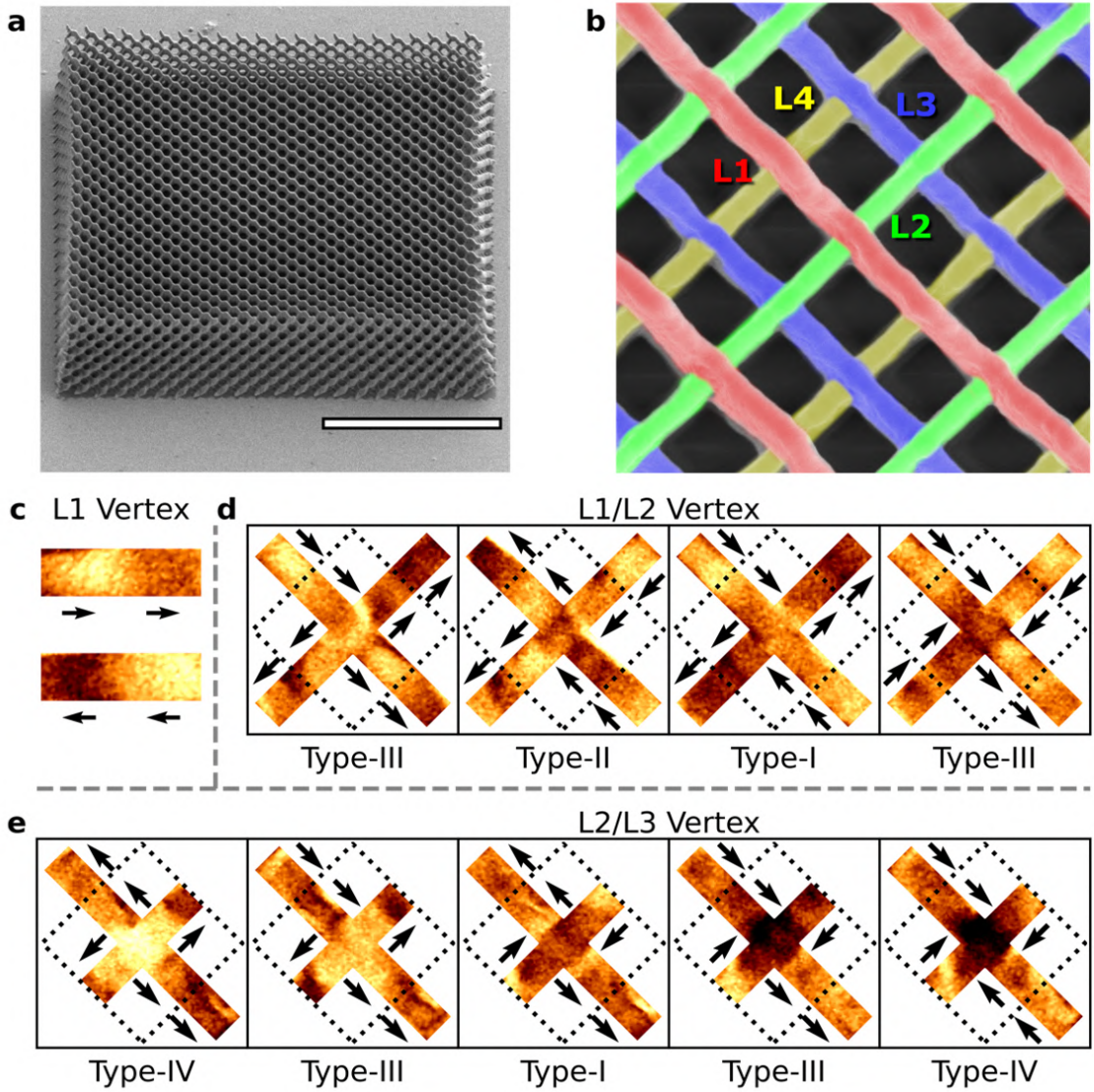

**Figure 2: Experimental vertex states and measured magnetic force microscopy contrast. (a)** A scanning electron microscope image of the 3D artificial spin-ice lattice, scale bar is 20 µm. **(b)** Top-view, false-colour SEM image of the 3DASI lattice with the individual sub-lattices labelled. Scale bar is 1 µm. **(c)** Magnetic force microscopy contrast of the vertex types found upon the L1 coordination two vertices. **(d)** Contrast for vertex types measured at the L1/L2 vertex. Here, both ice-rule (Type I, Type II) and single charged vertices (Type III) are observed. **(e)** Contrast for vertex types measured at the L2/L3 vertex, which shows a mixture of ice vertices as well as those with single and double magnetic charge (Type IV).

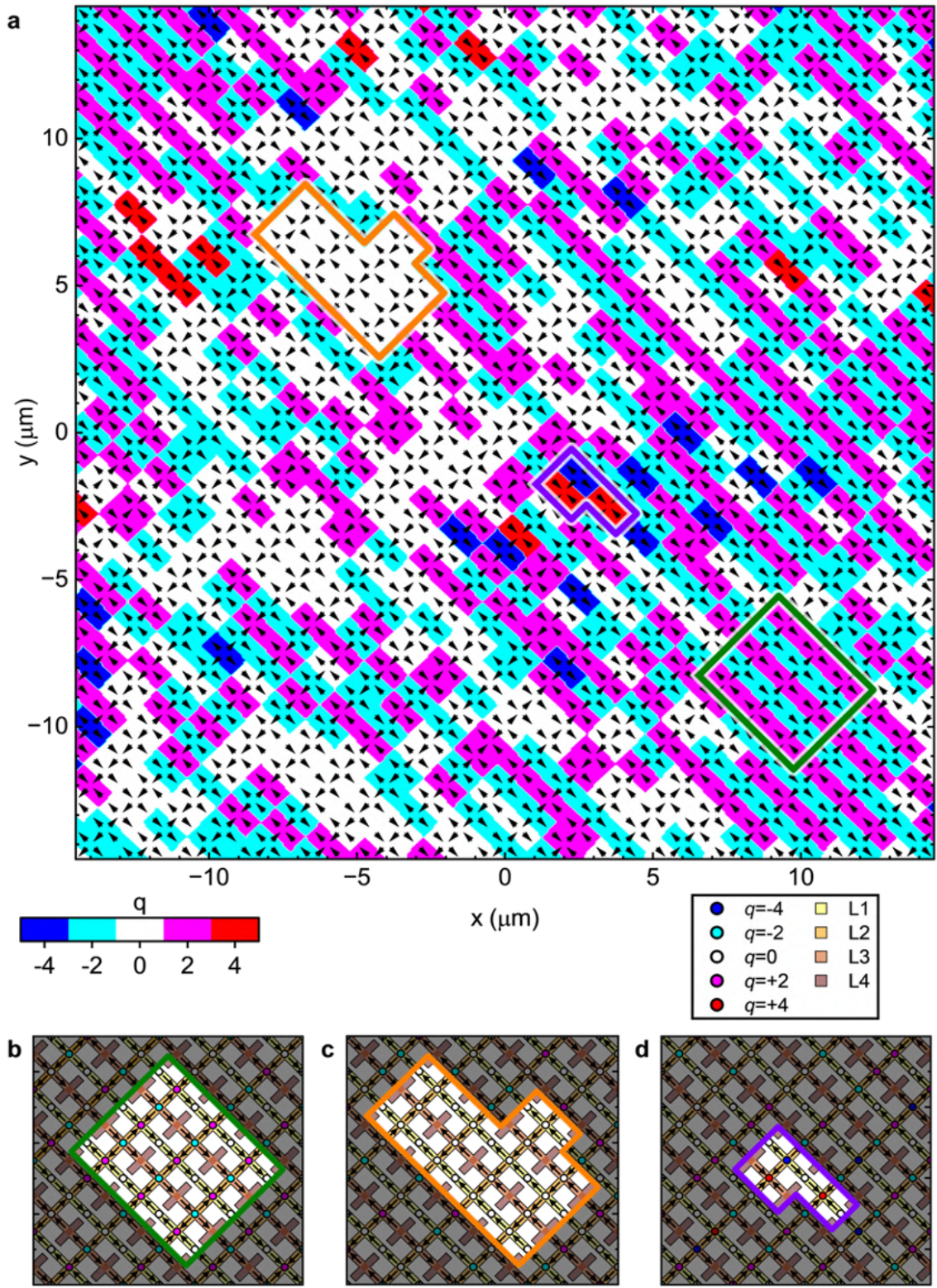

**Figure 3: Measuring the experimental demagnetised state of a 3D artificial spin-ice. (a)** Global magnetic charge map of the measured sample region. Charged regions are represented by colour as according to legend. The map shows examples of single charge crystallite (green outline), the ice phase (orange outline) and the double charge crystal (purple). **(b)** More detailed arrow map of the experimental single charge crystallite. It can be seen to consist of ferromagnetic stripes on the surface L1 layer with antiferromagnetic ordering upon L2. **(c)** Ice phase with type II tiling and **(d)** double charge crystallite, which only occurs at breaks in the L1 ferromagnetic ordering. Arrows represent the magnetisation of the L1, L2, and L3 sublattices. Here, the distinction between surface

coordination two, and sub-surface coordination four vertices can be seen. All sublattices are shaded to guide the eye. Magnetic charges are represented by circles of different colour, superimposed upon vertices.

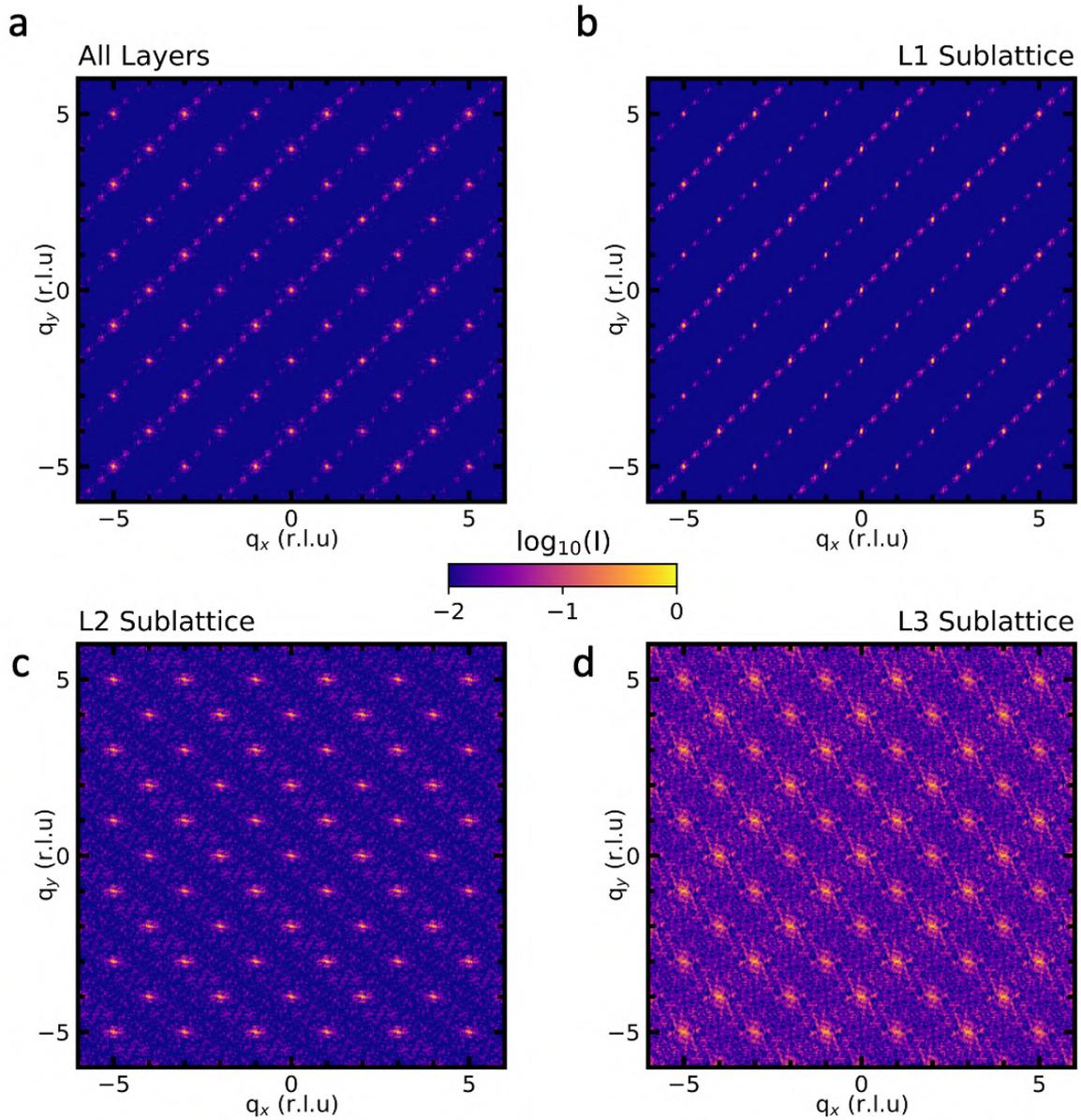

**Figure 4: Experimental magnetic structure factors. (a)** The magnetic structure factor of all sub-lattices superimposed. Clear Bragg peaks can be seen with periodicity in two dimensions. **(b)** The magnetic structure factor of the L1 sub-lattice. Peaks can be seen at q=[0,0], corresponding to the ferromagnetic ordering upon the surface. The split peaks about q=[1/2, 1/2], are due to domains on L1 with larger periodicity as demonstrated in Extended Data Fig 4. **(c)** The structure factor for the L2 sub-lattice. Bragg peaks can be seen, resulting from both type II vertices and charge crystallites **(d)** The structure factor for the L3 sub-lattice. Bragg peaks are again seen and match the periodicity seen for L2 sub-lattice, superimposed upon a diffuse background.

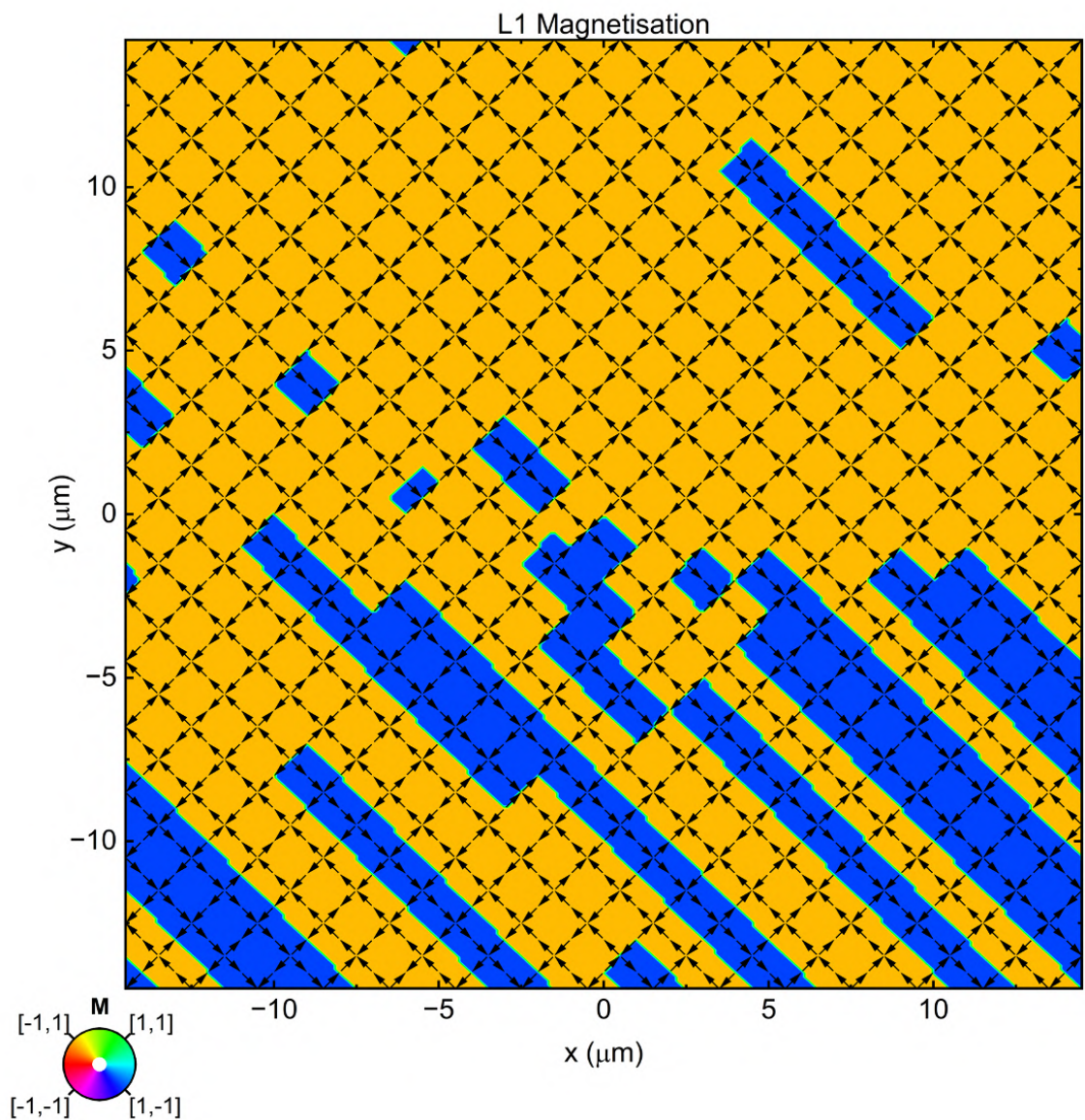

**Extended Data Figure 1**: Arrow map showing the magnetisation upon the L1 sub-lattice, which is seen to largely be composed of long ferromagnetic stripes.

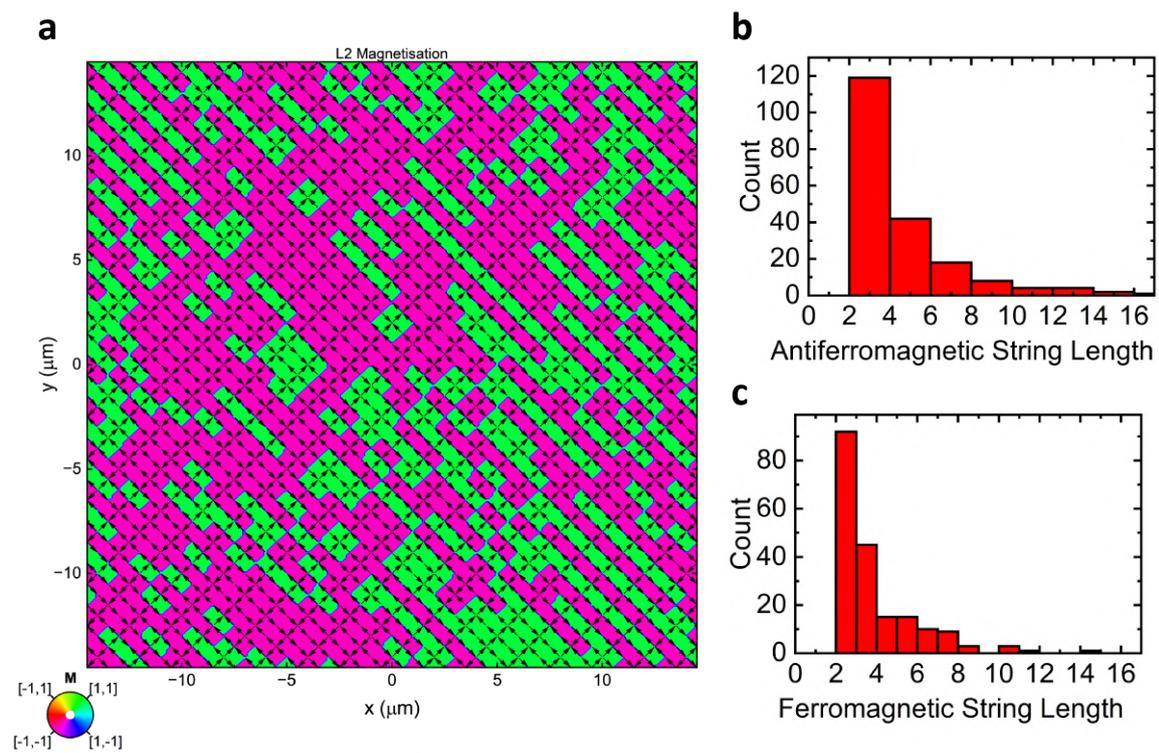

**Extended Data Figure 2**: (a) Arrow map showing the magnetisation upon the L2 sub-lattice. (b) Histogram of L2 antiferromagnetic string frequency. (c) Histogram of L2 ferromagnetic string frequency.

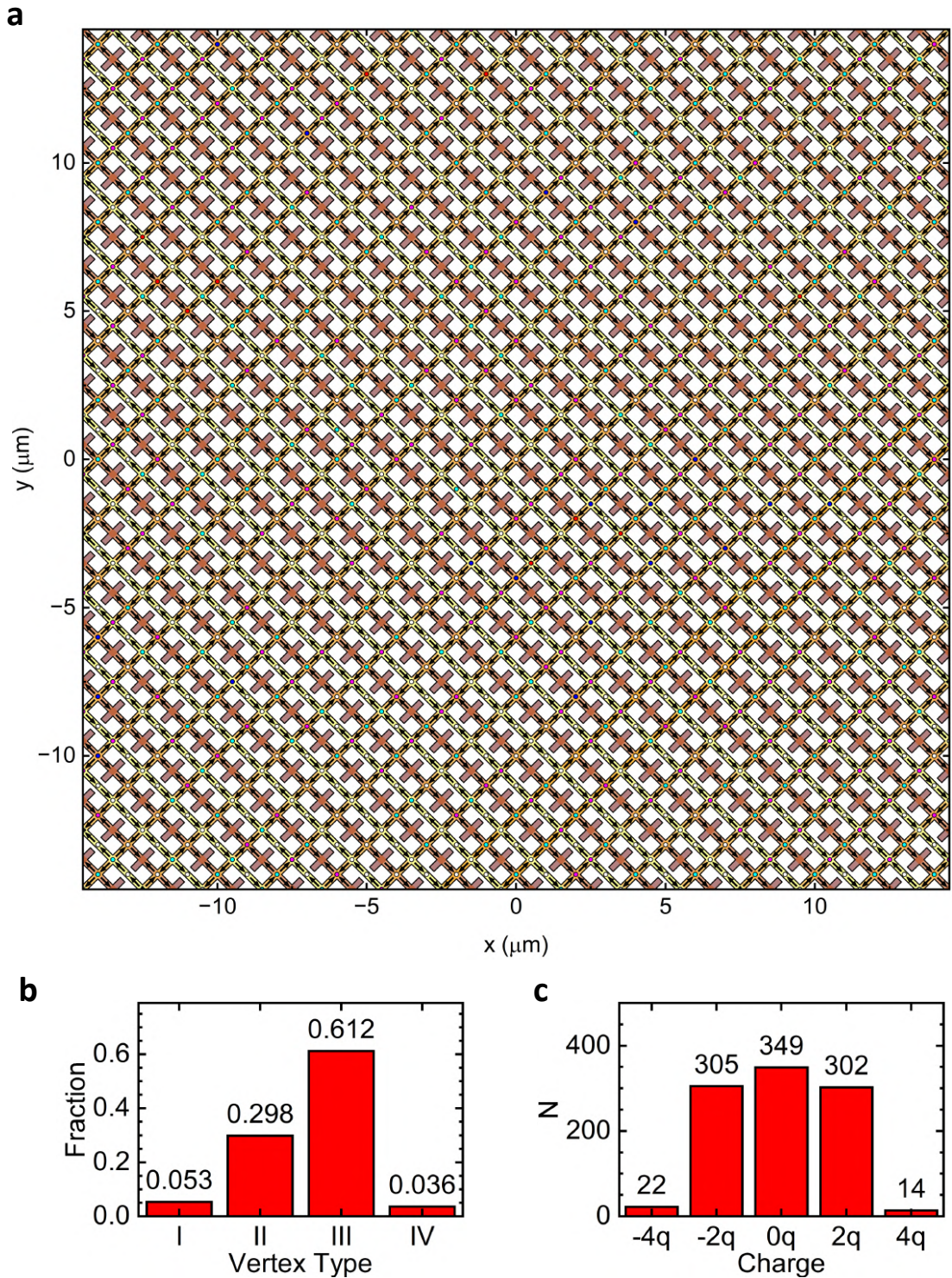

**Extended Data Figure 3:** (a) Full magnetisation map of the 30 μm x 30 μm sample area measured by magnetic force microscopy. (b) Vertex types as measured across the measured area. Coordination two vertices are not included. (c) Distribution of vertex charge across the measured area.

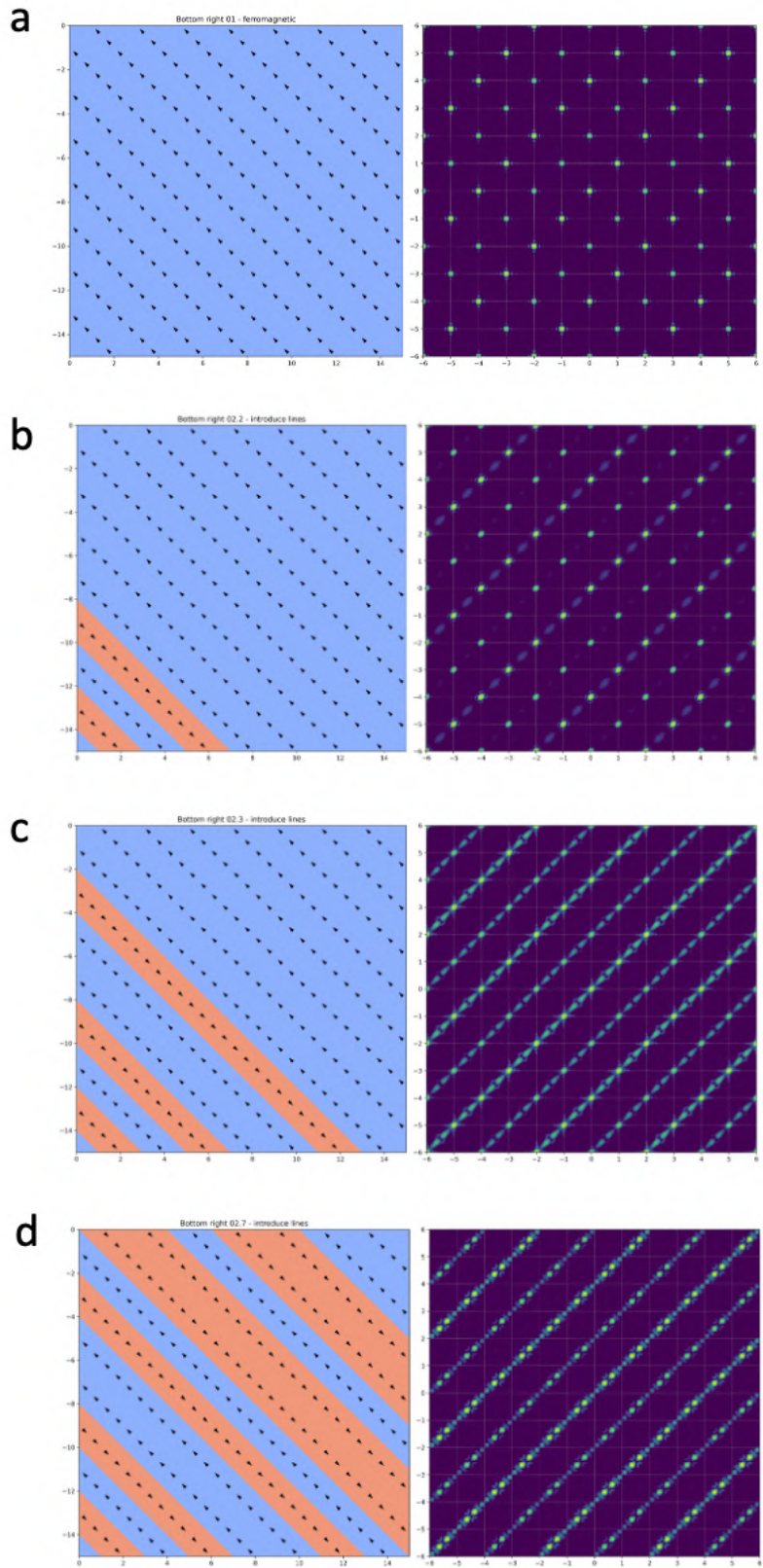

**Extended Data Figure 4:** Evolution of L1 structure factor, starting with **(a)** perfect ferromagnetic order and then **(b-d)** implementing domains seen in experimental arrow map.

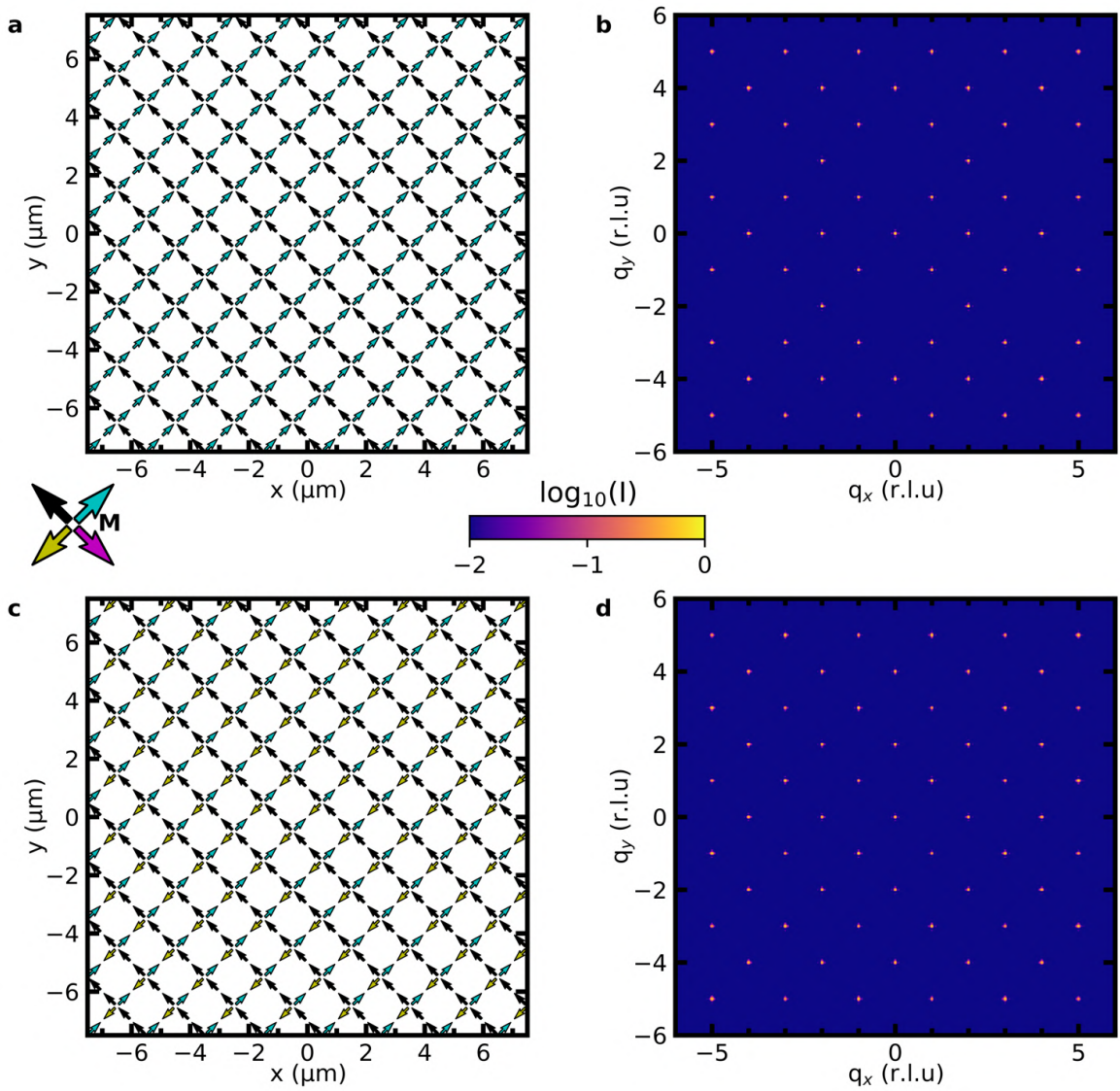

**Extended Data Figure 5:** Magnetic structure factor for (a) uniformly tiled type II vertices and for (b) a perfectly ordered experimental charge crystal ($C_EI$).

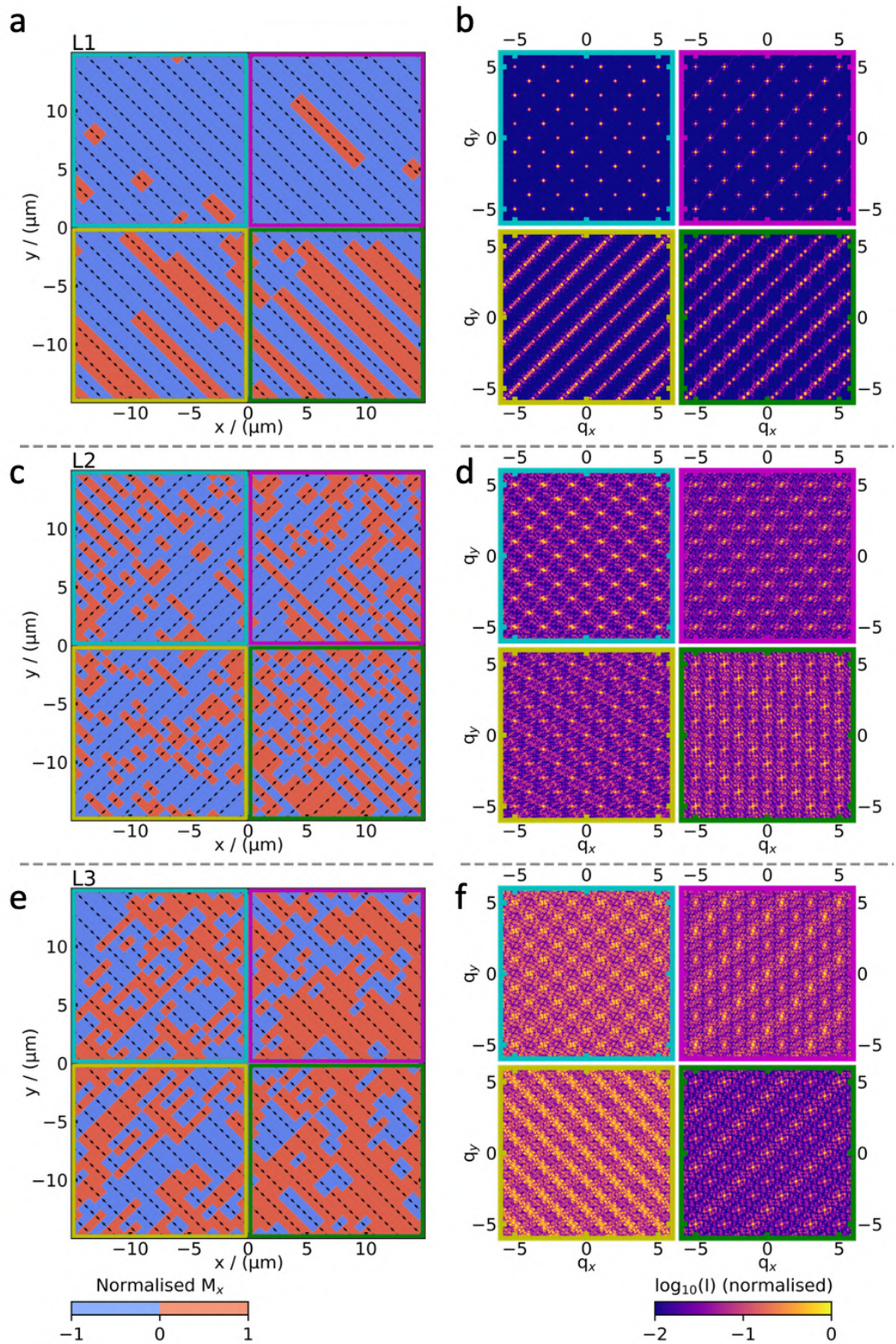

**Extended Data Figure 6:** Spin maps and associated structure factor for experimental data, separated by region and layer. (a) Arrow map for the L1 sub-lattice with (b) the associated structure factor. (c) Arrow map for the L2 sub-lattice with (d) the associated structure factor. (e) Arrow map for the L3 sub-lattice with (f) the associated structure factor. For each sub-lattice, regions as identified by square outlines have corresponding structure factors as shown in associated panel.

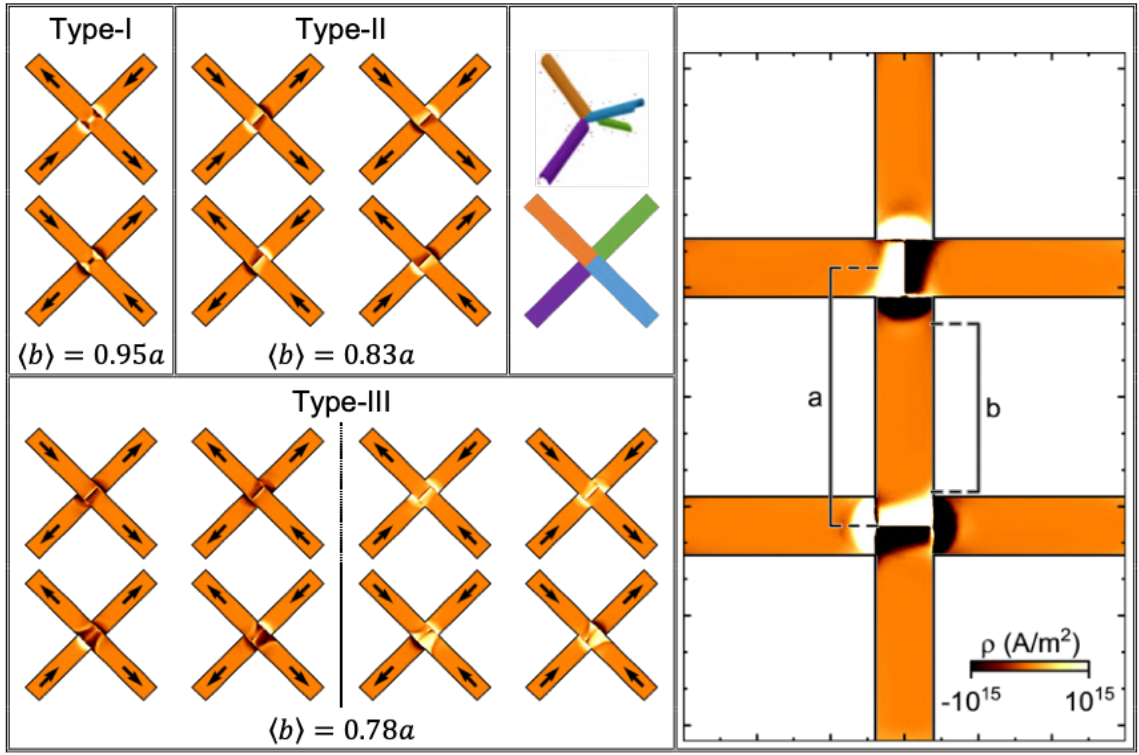

**Extended Data Figure 7: (a)** Magnetic charge density for vertex types computed using the NMag micromagnetic code. The lattice constant, a, is measured from vertex centre to vertex centre. The needle length, b, is measured across the region where the magnetisation is uniformly magnetised. Colour contrast shows the magnetic charge density as indicated in legend. **(b)** Table showing scaled dipolar needle length for range of vertices.

1. Klaus, L. *et al.* Observation of vortices and vortex stripes in a dipolar Bose-Einstein condensate. *Nature Physics* (2022). https://doi.org:https://doi.org/10.1038/s41567-022-01793-8
2. Ramirez, A. P., Hayashi, A., Cava, R. J., Siddharthan, R. & Shastry, B. S. Zero-point entropy in 'spin ice'. *Nature* **399**, 333-335 (1999).
3. Rosensweig, R. E. *Ferrohydrodynamics*. (Courier Corporation, 2013).
4. Tisza, J. M. and Tisza, L. Theory of Dipole Interaction in Crystals. *Phy. Rev.* **70**
5. Schildknecht, D., Schütt, M., Heyderman, L. J. & Derlet, P. M. Continuous ground-state degeneracy of classical dipoles on regular lattices. *Phys Rev B* **100**, 014426 (2019).
6. Bramwell, S. T. & Gingras, M. J. Spin ice state in frustrated magnetic pyrochlore materials. *Science* **294**, 1495-1501 (2001).
7. Castelnovo, C., Moessner, R., Sondhi, S. & Langer, J. Spin Ice, Fractionalization, and Topological Order. *Annual Review of Condensed Matter Physics, Vol 3* **3**, 35-55 (2012). https://doi.org:10.1146/annurev-conmatphys-020911-125058
8. Melko, R., den Hertog, B. & Gingras, M. Long-range order at low temperatures in dipolar spin ice. *Phys Rev Lett* **87**, 067203 (2001). https://doi.org:10.1103/PhysRevLett.87.067203
9. Ryzhkin, I. Magnetic relaxation in rare-earth oxide pyrochlores. *Journal of Experimental and Theoretical Physics* **101**, 481-486 (2005).
10. Castelnovo, C., Moessner, R. & Sondhi, S. L. Magnetic monopoles in spin ice. *Nature* **451**, 42-45 (2008).
11. Giblin, S. R., Bramwell, S. T., Holdsworth, P. C. W., Prabhakaran, D. & Terry, I. Creation and measurement of long-lived magnetic monopole currents in spin ice. *Nature Physics* **7**, 252-258 (2011). https://doi.org:Doi 10.1038/Nphys1896
12. Fennell, T. *et al.* Magnetic Coulomb phase in the spin ice $Ho_2Ti_2O_7$. *Science* **326**, 415-417 (2009).
13. Brooks-Bartlett, M., Banks, S., Jaubert, L., Harman-Clarke, A. & Holdsworth, P. Magnetic-Moment Fragmentation and Monopole Crystallization. *Phys Rev X* **4** (2014). https://doi.org:10.1103/PhysRevX.4.011007
14. Petit, S. *et al.* Observation of magnetic fragmentation in spin ice. *Nature Physics* **12**, 746-750 (2016).
15. Wang, R. F. *et al.* Artificial 'spin ice' in a geometrically frustrated lattice of nanoscale ferromagnetic islands (vol 439, pg 303, 2006). *Nature* **446**, 102-102 (2007). https://doi.org:Doi 10.1038/Nature05607
16. Skjærvø, S.H., Marrows, C.H, Stamps, R.L., Heyderman, L.J. Advances in artificial spin ice. *Nature Physics Reviews* **2**, 13-28 (2020).
17. Qi, Y., Brintlinger, T. & Cumings, J. Direct observation of the ice rule in an artificial kagome spin ice. *Phys Rev B* **77**, 094418 (2008). https://doi.org:10.1103/PhysRevB.77.094418
18. Drisko, J., Marsh, T. & Cumings, J. Topological frustration of artificial spin ice. *Nat Commun* **8**, 1-8 (2017).

19   Li, Y. *et al.* Superferromagnetism and domain-wall topologies in artificial "pinwheel" spin ice. *ACS Nano* **13**, 2213-2222 (2018).
20   Moller, G. & Moessner, R. Artificial square ice and related dipolar nanoarrays. *Phys Rev Lett* **96**, 237202 (2006).
21   Mol, L. A. S., Moura-Melo, W. A. & Pereira, A. R. Conditions for free magnetic monopoles in nanoscale square arrays of dipolar spin ice. *Phys Rev B* **82** (2010).
22   Perrin, Y., Canals, B. & Rougemaille, N. Extensive degeneracy, Coulomb phase and magnetic monopoles in artificial square ice. *Nature* **540**, 410-413 (2016). https://doi.org:10.1038/nature20155
23   Begum Popy, R., Frank, J. & Stamps, R. L. Magnetic field driven dynamics in twisted bilayer artificial spin ice at superlattice angles. *J Appl Phys* **132**, 133902 (2022).
24   May, A. *et al.* Magnetic charge propagation upon a 3D artificial spin-ice. *Nat Commun* **12** (2021). https://doi.org:10.1038/s41467-021-23480-7
25   Sahoo, S. et al. Observation of Coherent Spin Waves in a Three-Dimensional Artificial Spin Ice Structure. *Nano Lett* **21**, 4629-4635 (2021). https://doi.org:10.1021/acs.nanolett.1c00650
26   Sahoo, S. *et al.* Ultrafast magnetization dynamics in a nanoscale three-dimensional cobalt tetrapod structure. *Nanoscale* **10**, 9981-9986 (2018). https://doi.org:10.1039/c7nr07843a
27   Jaubert, L. D. C., Lin, T., Opel, T. S., Holdsworth, P. C. W. & Gingras, M. J. P. Spin ice Thin Film: Surface Ordering, Emergent Square ice, and Strain Effects. *Phys Rev Lett* **118** (2017). https://doi.org:ARTN 20720610.1103/PhysRevLett.118.207206
28   May, A., Hunt, M., Van Den Berg, A., Hejazi, A. & Ladak, S. Realisation of a frustrated 3D magnetic nanowire lattice. *Communications Physics* **2**, 1-9 (2019).
29   Wang, Y.L. *et al.* Rewritable artificial magnetic charge ice. *Science* **352**, 962-966 (2016).
30   Zhang, S. *et al.* Crystallites of magnetic charges in artificial spin ice. *Nature* **500**, 553-557 (2013). https://doi.org:10.1038/nature12399
31   Yue, W.C. *et al.* Crystallizing Kagome artificial spin ice. *Phys Rev Lett* **129**, 057202 (2022).
32   Hofhuis, K. *et al.* Real-space imaging of phase transitions in bridged artificial kagome spin ice. *Nature Physics*, 1-7 (2022).
33   Rougemaille, N. & Canals, B. The magnetic structure factor of the square ice: A phenomenological description. *Appl Phys Lett* **118**, 112403 (2021).
34   Wang, R. F. *et al.* Demagnetization protocols for frustrated interacting nanomagnet arrays. *Journal of Applied Physics* **101**, 09J104 (2007). https://doi.org:10.1063/1.2712528
35   Petai Pip et al. X-ray imaging of the magnetic configuration of a three-dimensional artificial spin ice building block. *APL Materials* **10**, 101101 (2022)